\documentclass[12pt]{iopart}

\usepackage{iopams}
\usepackage{calrsfs}
\usepackage{graphicx}
\begin{document}

\title[]{Bose-Einstein condensation of heteronuclear bound states formed in a Fermi gas of two atomic species: microscopic approach}

\author{A S Peletminskii$^1$, S V Peletminskii$^1$ and Yu V Slyusarenko$^{1,2}$}

\address{$^1$ National Science Center
"Kharkov Institute of Physics and Technology", 1, Akademicheskaya Str., 61108 Kharkov, Ukraine}
\address{$^2$ Karazin National University, 4, Svobody Sq., 61077 Kharkov, Ukraine}

\ead{aspelet@kipt.kharkov.ua}
\vspace{10pt}

\begin{abstract}
We study a many-body system of interacting fermionic atoms of two species that are in thermodynamic equilibrium with their condensed heteronuclear bound states (molecules). In order to describe such an equilibrium state, we use a microscopic approach that involves the Bogoliubov model for a weakly interacting Bose gas and approximate formulation of the second quantization method in the presence of bound states of particles elaborated earlier by the authors. This microscopic approach is valid at low temperatures, when the average kinetic energy of all the components in the system is small in comparison with the bound state energy. The coupled equations, which relate the chemical potentials of fermionic components and molecular condensate density, are obtained within the proposed theory. At zero temperature, these equations are analyzed both analytically and numerically, attracting the relevant experimental data. We find the conditions at which a condensate of heteronuclear molecules coexists in equilibrium with degenerate components of a Fermi gas. The ground state energy and single-particle excitation spectrum are found. The boundaries of the applicability of the developed microscopic approach are analyzed.
\end{abstract}

\pacs{05.30.Fk, 05.30.Jp, 03.75.Hh, 67.85.Pq}
%
\vspace{2pc}
\noindent{\it Keywords}: ultracold gases, mixtures of bosons and fermions, heteronuclear bound states, Bose-Einsteum condensation
%
%
%
%

\section{Introduction}

The study of the formation of diatomic molecules in ultracold dilute quantum gases of bosonic and fermionic atoms has attracted much interest over the past years. Such molecules, being weakly bound states of two atoms, can be produced through the single-photon photoassociation of condensed atoms \cite{McKenzie} or by two-photon stimulated Raman transition \cite{Wynar}. However, a Feshbach resonance has been found to be the most powerful tool for controlling the interaction between atoms and creating the diatomic molecules \cite{Kohler,Grimm}.

The first experiments to produce the diatomic molecules were performed in dilute single-species atomic gases. In particular, the homonuclear molecules ${}^{23}{\rm Na}_{2}$ \cite{Xu} and ${}^{133}{\rm Cs}_{2}$ \cite{Herbig} formed from cold bosonic atoms were produced. These molecules are in highly vibrationally excited states and usually undergo a fast decay over $10^{-4}$ s, whereas the molecules ${}^{6}{\rm Li}_{2}$ \cite{Strecker,Cubizolles,Jochim}, created from fermionic atoms in different internal states, exhibits a sufficiently long lifetime, of the order of 1 s. As has been shown \cite{Shlyapnikov,Petrov}, the long lifetime of molecules consisting of fermionic atoms is a consequence of the manifestation of Fermi statistics in the interaction of atoms that form a molecule.

Recently, interest has focused on the creation of molecules consisting of atoms of different atomic species, or the so-called heteronuclear molecules. In particular, the Bose-Bose \cite{Papp,Weber,Dutta,Koppinger} and Bose-Fermi \cite{Ospelkaus} molecules have been
experimentally realized. However, Fermi-Fermi heteronuclear molecules have attracted the most interest, since  they, as we have already mentioned, are expected to
exhibit a sufficiently long lifetime \cite{Shlyapnikov,Petrov}. The first such molecules with a lifetime more than 100 ms were created in a dilute mixture of ${}^{6}{\rm Li}$ and ${}^{40}{\rm
K}$ atoms \cite{Voigt}. In addition, the simultaneous quantum degeneracy realized in a gas mixture consisting of ${}^{6}{\rm Li}$ and ${}^{173}{\rm Yb}$ fermionic atoms gives the possibility of producing another kind of heteronuclear Fermi-Fermi molecules \cite{Hara}. The long lifetime of molecules consisting of fermionic atoms is a good basis for achieving molecular Bose-Einstein condensation.

Theoretical descriptions of degenerate ultracold gases interacting through a Feshbach resonance have been presented both for bosonic \cite{Drumond,Timmermans1,Timmermans2,Annals} and fermionic \cite{Andreev,Gurarie,Sheehy} atoms. However, the resonant fermionic gases have attracted much more interest, since they demonstrate different regimes of a superfluid (superconducting) phenomenon: condensation of Copper pairs (BCS state), Bose-Einstein condensation of molecules (BEC state) \cite{Eagles,Leggett,Nozieres}, and BEC-BCS crossover \cite{Giorgini,Zwerger}.

When we want to describe a many-body system of interacting fermionic atoms and their diatomic bound states at the microscopic level, the inevitable question arises as to in what way the characteristics of interactions of bound states are related to those of original fermionic atoms. Such a problem has been studied on the basis of the Schr\"{o}dinger equation in \cite{Shlyapnikov,Petrov}, where the authors managed to express the dimer-dimer scattering length in terms of the scattering length of atoms with opposite spins. If such dimers can be regarded as molecules, then the latter are extremely loose, because the binding energy of dimers is very small and their size is close to the scattering length of atoms.

Another approach to the problem, which develops an approximate second quantization method in the presence of bound states of two different fermions has been studied in \cite{Pelet-Slyus}. The correct construction of such formulation of the second quantization method is possible if the average kinetic energy of particles is small in comparison with the energy of their bound states. The key problem of the developed formalism is to introduce, in an appropriate way, the creation and annihilation operators of two-fermion bound state as the whole object and to preserve, at the same time, the information regarding its quantum states. It is clear that in this case the size of the bound state should be small compared with the scattering length. The elaborated method has enabled the second quantized pair-interaction Hamiltonian to be obtained, in which all the interaction amplitudes (or coupling constants) are expressed in terms of the interaction amplitudes of original fermions, which form the bound states.

The results of the formulated approximate second quantization method have been tested to study quantum-electrodynamic processes such as spontaneous emission of an atom, scattering of photons and fermions by an atom, as well as to derive the expression for the Wan der Waals forces
\cite{Pelet-Slyus}. They have also found application in the physics of ultracold gases, in particular when studying slowing-down of light in an atomic Bose-Einstein condensate \cite{Slyus,Sotn}.

The present paper examines the Bose-Einstein condensation of  'stable' diatomic bound states or 'molecules' formed from the fermionic atoms of two different species. To the best of the authors' knowledge, the formulation of the problem itself dealing with existence of Bose condensed phase in such a system is novel. The stability of the bound states is guaranteed by the smallness of the kinetic energy of atoms in comparison to the energy of their bound states. In the following, we use the term 'bound states' to avoid the consideration of the structure of the real molecular spectrum that has a quite complex structure. In this sense, the bound states represent the simplest model of a molecule. It is worth stressing that we do not study a mechanism responsible for the production of diatomic bound states but assume that, at certain conditions, they are in thermodynamic equilibrium with unbound fermionic atoms. A starting point of our work is the microscopic Hamiltonian described above \cite{Pelet-Slyus} that specifies the interactions of fermionic atoms with their bound states and between the bound states themselves.
Note that the interaction between all the components of the system should be weak enough to ensure the stability of bound states in equilibrium (see section 4). This fact allows us to apply the Bogoliubov microscopic model for a weakly interacting Bose gas \cite{Bogoliubov} to describe a condensate of bound states coupled to unbound fermionic atoms. However, as we shall see below, the standard Bogoliubov model requires significant modification for such a specific system.

\section{Microscopic second quantized Hamiltonian for two species of fermions and their heteronuclear bound states}

Before starting to solve the declared problem, let us sketch out the basic aspects of the above-mentioned formulation of an approximate second quantization method in the presence of bound states of particles.
It is well known that the second quantization method is a powerful tool that is usually used to describe the physical processes in quantum many-body systems. The particle creation and annihilation operators are the key concept of this method since the operators of relevant physical quantities are constructed in terms of them. Such a description is absolutely accurate at arbitrary interaction between the particles and implies that the particles are elementary objects, i.e. not consisting of other particles.

However, the interparticle interaction may lead to the formation of bound states. In this case, the standard exact formulation (in the sense mentioned above) of the second quantization method becomes too cumbersome. Therefore, one might expect that a consistent quantum-mechanical theory of many-body systems should involve consideration of the 'elementary' particles as well as the possible existence of their bound states. Moreover, in this theory, it is necessary to define the creation and annihilation operators of bound states as operators of elementary (not compound) objects and to preserve, at the same time, the necessary information regarding the internal degrees of freedom of bound states. It is clear that such a formulation of the second quantization method is approximate in contrast to the standard exact formulation. In particular, it can be realized when the bound state energy of compound particles is great compared to the kinetic energy of original particles and results in the following second quantized Hamiltonian \cite{Pelet-Slyus}:
\begin{equation}
H=H_{0}+V, \label{eq:1.1}
\end{equation}
where $H_{0}$ is the kinetic energy operator,
\begin{eqnarray}
H_{0}&=\sum_{i=1}^{2}{1\over 2m_{i}} \int d{\bf x}{\partial\chi_{i}^{\dagger}({\bf x}) \over\partial{\bf x}}
{\partial\chi_{i}({\bf x}) \over\partial{\bf x}} \nonumber \\
&+\sum_{\alpha}\int d{\bf X}\biggl[{1\over 2M}{\partial\eta_{\alpha}^{\dagger}({\bf X}) \over\partial{\bf X}}{\partial\eta_{\alpha}({\bf
X}) \over\partial{\bf X}}+\varepsilon_{\alpha}\eta_{\alpha}^{\dagger}({\bf X})\eta_{\alpha}({\bf X})\biggr]. \label{eq:1.2}
\end{eqnarray}
Here $m_{i}$ $(i=1,2)$ is the mass of a fermionic atom of the first or second species, $\chi_{i}^{\dagger}({\bf x})$ and $\chi_{i}({\bf x})$ are the creation and annihilation operators, respectively, of these atoms at the point ${\bf x}$, $M=m_{1}+m_{2}$ is the mass of a bound state (the bound states are formed from the interspecies fermionic atoms), $\eta_{\alpha}^{\dagger}({\bf X})$, $\eta_{\alpha}({\bf X})$ are the field creation and annihilation operators of diatomic bound states at the point with a coordinate of the center of mass ${\bf X}$. The energy spectrum of a bound state $\varepsilon_{\alpha}$ is found from the Schr\"{o}dinger equation
\begin{equation} \label{eq:1.3}
\biggl[-{1\over 2m_{*}}\Delta+\nu_{12}({\bf
x})\biggl]\phi_{\alpha}({\bf
x})={\varepsilon}_{\alpha}\phi_{\alpha}({\bf x}),
\end{equation}
where $m_{*}={m_{1}m_{2}/(m_{1}+m_{2})}$ is the reduced mass, $\nu_{12}({\bf x})$ is the interaction amplitude between the fermionic atoms of the first and second species, and $\phi_{\alpha}({\bf x})$ is the wave function of the bound state. The Greek index '$\alpha$'\, is used to denote the whole set of the quantum numbers, which specify a quantum-mechanical 'molecular'\, state. The wave functions are assumed to satisfy the orthonormality condition,
\begin{equation}\label{eq:1.4}
\int d{\bf x}\, \phi_{\alpha}({\bf x})\phi^{*}_{\beta}({\bf x})=\delta_{\alpha\beta}.
\end{equation}
The interaction Hamiltonian in Eq. (\ref{eq:1.1}) is represented as $V=V_{bb}+V_{ff}+V_{bf}$ with $V_{bb}$ being the interaction between the bound states,
\begin{eqnarray}
V_{bb}&={1\over 2}\int d{\bf x}_{1}\int d{\bf x}_{2} \int d{\bf y}_{1}\int d{\bf y}_{2} \varphi^{\dagger}({\bf x}_{1},{\bf
y}_{1})\varphi^{\dagger}({\bf x}_{2},{\bf y}_{2}) \varphi({\bf x}_{2},{\bf y}_{2})\varphi({\bf x}_{1},{\bf
y}_{1}) \nonumber \\
&\times\biggl[\nu_{11}({\bf x}_{1}-{\bf x}_{2})+\nu_{22}({\bf y}_{1}-{\bf y}_{2})+\nu_{12}({\bf x}_{1}-{\bf y}_{2})+ \nu_{21}({\bf y}_{1}-{\bf
x}_{2}) \biggr], \label{eq:1.5}
\end{eqnarray}
where $\nu_{ij}({\bf x})$ are the interaction amplitudes of unbound fermionic atoms of the species $i$ and $j$ ($i,j=1,2$) and
\begin{equation} \label{eq:1.6}
\varphi({\bf x},{\bf y})\equiv \sum_{\alpha} \phi_{\alpha}({\bf x}-{\bf y}){\eta}_{\alpha}({\bf X}), \quad {\bf X}={m_{1}{\bf x}+m_{2}{\bf
y}\over m_{1}+m_{2}}
\end{equation}
is different from zero when ${\bf x}\approx {\bf y}$. The interaction of bound states with fermionic atoms of both species is specified by the following operator:
\begin{eqnarray}
\fl V_{bf}=\int d{\bf x}_{1}\int d{\bf x}_{2}\int d{\bf y}_{2}\varphi^{\dagger}({\bf x}_{2},{\bf y}_{2})\varphi({\bf x}_{2}, {\bf y}_{2}) \biggl[\left[\nu_{11}({\bf x}_{1}-{\bf x}_{2})+
\nu_{21}({\bf x}_{1}-{\bf y}_{2})\right]\chi_{1}^{\dagger}({\bf x}_{1})\chi_{1}({\bf x}_{1}) \nonumber  \\
+[\nu_{22}({\bf x}_{1}-{\bf y}_{2})+\nu_{12}({\bf x}_{1}-{\bf x}_{2})] \chi_{2}^{\dagger}({\bf x}_{1}) \chi_{2}({\bf x}_{1})\biggr].
\label{eq:1.7}
\end{eqnarray}
Finally, $V_{ff}$, which characterizes the interaction between unbound atoms themselves, has the form
\begin{eqnarray}
V_{ff}&={1\over 2}\int d{\bf x}_{1}\int d{\bf x}_{2}\biggl[\nu_{11}({\bf x}_{1}-{\bf x}_{2})\chi_{1}^{\dagger}({\bf x}_{1}) \chi_{1}^{\dagger}({\bf
x}_{2})\chi_{1}({\bf x}_{2})\chi_{1}({\bf x}_{1}) \nonumber \\
&+\nu_{22}({\bf x}_{1}-{\bf x}_{2})\chi_{2}^{\dagger}({\bf x}_{1})  \chi_{2}^{\dagger}({\bf
x}_{2})\chi_{2}({\bf x}_{2})\chi_{2}({\bf x}_{1}) \nonumber \\
&+2\nu_{12}({\bf x}_{1}-{\bf x}_{2})\chi_{1}^{\dagger}({\bf x}_{1}) \chi_{1}({\bf x}_{1})\chi_{2}^{\dagger}({\bf x}_{2}) \chi_{2}({\bf x}_{2})\biggr].
\label{eq:1.8}
\end{eqnarray}

Thus, all the interactions in which the bound states are involved are expressed in terms of the interaction amplitudes $\nu_{ij}({\bf x})$ of original unbound fermionic atoms. Moreover, the interaction Hamiltonian does not take into account the terms responsible for the processes associated with conversion of atoms into a bound state, and its reconversion to unbound atoms as it should be in the low-energy approximation. Therefore, the bound states are absolutely stable in the leading approximation.

Below, we study Bose-Einstein condensation of heteronuclear bound states. Therefore, it is convenient to write the Hamiltonian in the momentum representation, since, according to Bogoliubov's method \cite{Bogoliubov}, one needs to extract the condensate amplitudes, i.e. to replace the corresponding bosonic creation and annihilation operators with zero momentum by $c$ -- numbers. To this end, let us introduce the creation $a^{\dagger}_{{i\bf p}}$ and annihilation $a_{{i\bf p}}$ operators of fermionic atoms (index $i$ denotes the atomic species) along with creation $b_{\alpha{\bf p}}^{\dagger}$ and annihilation $b_{\alpha{\bf p}}$ operators of their diatomic bound states,
\begin{eqnarray}
\chi_{i}({\bf x})={1\over\sqrt{\mathcal{V}}}\sum_{{\bf p}}e^{i{\bf qx}/\hbar}a_{i{\bf p}}, \quad i=1,2, \nonumber \\
\varphi({\bf x}_{1},{\bf y}_{1})={1\over\sqrt{\mathcal{V}}}\sum_{{\bf p},\alpha}e^{i{\bf pX}/\hbar}\phi_{\alpha}({\bf x}_{1}-{\bf y}_{1})b_{\alpha{\bf p}}, \label{eq:1.9}
\end{eqnarray}
where ${\bf X}=(m_{1}{\bf x}_{1}+m_{2}{\bf y}_{1})/(m_{1}+m_{2})$ is the center of mass coordinate and $\mathcal{V}$ is the volume of the
system. Then the kinetic energy operator given by Eq. (\ref{eq:1.2}) reads
\begin{equation} \label{eq:1.10}
H_{0}=\sum_{i=1}^{2}\sum_{{\bf p}}{p^{2}\over 2m_{i}}a^{\dagger}_{i{\bf p}}a_{i{\bf p}}+\sum_{{\bf p},\alpha}\left[{p^{2}\over
2M}+\varepsilon_{\alpha}\right]b^{\dagger}_{{\bf p}\alpha}b_{{\bf p}\alpha}.
\end{equation}
The interaction Hamiltonian of bound states, according to Eq. (\ref{eq:1.5}), takes the form
\begin{equation}\label{eq:1.11}
V_{bb}={1\over 2\mathcal{V}}\sum_{{\bf p}_{1}...{\bf p_{4}}}g_{\alpha\delta\gamma\beta}({\bf p}_{14})\,b^{\dagger}_{{\bf
p}_{1}\alpha}b^{\dagger}_{{\bf p}_{2}\beta}b_{{\bf p}_{3}\gamma}b_{{\bf p}_{4}\delta}\,\delta_{{\bf p}_{1}+{\bf p}_{2},{\bf p}_{3}+{\bf p}_{4}},
\end{equation}
where
\begin{eqnarray}
g_{\alpha\delta\gamma\beta}({\bf p}_{14})=\sigma_{\alpha\delta}^{(2)}({\bf p}_{14})\nu_{11}({\bf p}_{14})\sigma_{\gamma\beta}^{*(2)}({\bf p}_{14}) +\sigma_{\delta\alpha}^{*(1)}({\bf p}_{14})\nu_{22}({\bf p}_{14})\sigma_{\beta\gamma}^{(1)}({\bf p}_{14}) \nonumber  \\
+\sigma_{\alpha\delta}^{(2)}({\bf p}_{14})\nu_{12}({\bf p}_{14})\sigma_{\beta\gamma}^{(1)}({\bf p}_{14})+\sigma_{\delta\alpha}^{*(1)}({\bf
p}_{14})\nu_{21}({\bf p}_{14})\sigma_{\gamma\beta}^{*(2)}({\bf p}_{14}) \label{eq:1.12}
\end{eqnarray}
and
\begin{equation}\label{eq:1.13}
\sigma_{\alpha\beta}^{(i)}({\bf p})=\int d{\bf x}\,\phi_{\alpha}^{*}({\bf x})\phi_{\beta}({\bf x})\exp\left[ i{{\bf p}{\bf x}}{m_{i}\over
\hbar M}\right], \quad i=1,2.
\end{equation}
Here ${\bf p}_{ij}\equiv {\bf p}_{i}-{\bf p}_{j}$ and $\nu_{ij}({\bf p}_{ij})=\nu_{ij}({\bf p}_{ji})$.
In Eq. (\ref{eq:1.11}) and below we assume the summation over the repeated Greek indices, which denote the quantum-mechanical 'molecular' state. Since the interaction amplitude $g_{\alpha\delta\gamma\beta}({\bf p}_{14})$ meets the relationship $g_{\alpha\delta\gamma\beta}({\bf p}_{14})=g^{*}_{\delta\alpha\beta\gamma}({\bf p}_{41})$, which is valid due to the evident property $\sigma_{\alpha\beta}^{(i)}({\bf p})=\sigma_{\beta\alpha}^{*(i)}(-{\bf p})$, $V_{bb}$ is the Hermitian operator. The interaction of bound states with unbound fermionic atoms, given by Eq. (\ref{eq:1.7}), is also expressed through $\sigma_{\alpha\beta}^{(i)}({\bf p})$,
\begin{eqnarray}
V_{bf}&={1\over\mathcal{V}}\sum_{{\bf p}_{1}...{\bf p_{4}}}v_{\alpha\beta}({\bf p}_{21})\,b^{\dagger}_{{\bf p}_{1}\alpha}b_{{\bf p}_{2}\beta}a_{1{\bf p}_{3}}^{\dagger}a_{1{\bf p}_{4}}\,\delta_{{\bf p}_{2}-{\bf p}_{1},{\bf p}_{3}-{\bf p}_{4}} \nonumber \\
&+{1\over \mathcal{V}}\sum_{{\bf p}_{1}...{\bf p_{4}}}u_{\alpha\beta}({\bf p}_{21})\,b^{\dagger}_{{\bf p}_{1}\alpha}b_{{\bf
p}_{2}\beta}a_{2{\bf p}_{3}}^{\dagger}a_{2{\bf p}_{4}}\,\delta_{{\bf p}_{2}-{\bf p}_{1},{\bf p}_{3}-{\bf p}_{4}}, \label{eq:1.14}
\end{eqnarray}
where
\begin{eqnarray}
v_{\alpha\beta}({\bf p}_{21})=\nu_{11}({\bf p}_{21})\sigma_{\beta\alpha}^{*(2)}({\bf p}_{21})+
\nu_{21}({\bf p}_{21})\sigma_{\alpha\beta}^{(1)}({\bf p}_{21}),\nonumber \\
u_{\alpha\beta}({\bf p}_{21})=\nu_{22}({\bf p}_{21})\sigma_{\alpha\beta}^{(1)}({\bf p}_{21})+ \nu_{12}({\bf
p}_{21})\sigma_{\beta\alpha}^{*(2)}({\bf p}_{21}). \label{eq:1.15}
\end{eqnarray}
Using the above property of $\sigma_{\alpha\beta}^{(i)}({\bf p})$, one can show that $V_{bf}$ is also the Hermitian operator. Finally, the interaction Hamiltonian of fermionic atoms of both species, according to Eq. (\ref{eq:1.8}), is given by
\begin{eqnarray}
V_{ff}&={1\over 2\mathcal{V}}\sum_{{\bf p}_{1}...{\bf p_{4}}}\nu_{11}({{\bf p}_{14}})\,a^{\dagger}_{1{\bf p}_{1}}a^{\dagger}_{1{\bf p}_{2}}a_{1{\bf p}_{3}}a_{1{\bf p}_{4}}\,\delta_{{\bf p}_{1}+{\bf p}_{2},{\bf p}_{3}+{\bf p}_{4}} \nonumber \\
&+{1\over 2\mathcal{V}}\sum_{{\bf p}_{1}...{\bf p_{4}}}\nu_{22}({{\bf p}_{14}})\,a^{\dagger}_{2{\bf p}_{1}}a^{\dagger}_{2{\bf p}_{2}}a_{2{\bf p}_{3}}a_{2{\bf p}_{4}}\,\delta_{{\bf p}_{1}+{\bf p}_{2},{\bf p}_{3}+{\bf p}_{4}} \nonumber \\
&+{1\over \mathcal{V}}\sum_{{\bf p}_{1}...{\bf p_{4}}}\nu_{12}({{\bf p}_{12}})\,a^{\dagger}_{1{\bf p}_{1}}a_{1{\bf p}_{2}}a_{2{\bf
p}_{3}}^{\dagger}a_{2{\bf p}_{4}}\,\delta_{{\bf p}_{1}-{\bf p}_{2},{\bf p}_{4}-{\bf p}_{3}}. \label{eq:1.16}
\end{eqnarray}

To conclude this section, note that the necessity to take into account both compound and elementary particles in the system is a typical problem, which occurs, e.g., when studying the interaction of radiation with matter consisting of neutral atoms or molecules in the ground or excited states. In this case, one needs to take into account the internal structure of atoms and molecules preserving, at the same time, the convenience and simplicity of the second quantization method. Such a situation also occurs when attempting to describe the experiments on laser cooling of atoms or when studying the chemical reactions in quantum chemistry. Below, we apply the above Hamiltonian to examine Bose-Einstein condensation of diatomic heteronuclear bound states, which are formed from interspecies fermionic atoms. These atoms are assumed to be in the given spin states so that we do not consider the interactions that affect the spin degrees of freedom.

\section{Generalized Bogoliubov model}

Consider a many-body system of interacting two-species fermionic atoms. Suppose that the interaction between them leads to the formation of heteronuclear diatomic bound states, which are Bose condensed at ultralow temperatures. It is worth stressing that we do not study a mechanism responsible for the formation of bound states but assume that at certain conditions the existing bound states are in thermodynamic equilibrium with unbound fermionic atoms. In order to describe such a system, we address the Hamiltonian given by Eqs. (\ref{eq:1.10})-(\ref{eq:1.16}). As has been already noted, it is obtained from the microscopic pair-interaction Hamiltonian of original two-species fermionic atoms and is expressed through the amplitudes $\nu_{ij}({\bf p})$ of their interaction. In experiments on the creation of ultracold heteronuclear molecules, the fermionic atoms of both species are prepared in pure quantum states $|1\rangle$ and $|2\rangle$, respectively. For example, in the case of ${}^{6}{\rm Li}$--${}^{40}{\rm
K}$ molecules, the states of fermionic atoms are K$|F=9/2,m_{F}=-5/2\rangle$ and Li$|F=1/2,m_{F}=1/2\rangle$, where $F$ and $m_{F}$ are the total spin and its projection \cite{Voigt}. Therefore, we assume that the fermionic atoms are characterized by the energy levels $\varepsilon_{1}$ and $\varepsilon_{2}$. Their bound states are also in a given internal state with energy $\varepsilon$. Then the Hamiltonian represented in the preceding section is reduced to a more simple form,
\begin{equation} \label{eq:2.1}
H=H_{0}+V,
\end{equation}
where
\begin{equation} \label{eq:2.2}
H_{0}=\sum_{{\bf p}}\left[{p^{2}\over 2M}+\varepsilon\right]b^{\dagger}_{\bf p}b_{\bf p}+\sum_{i,{\bf p}}\left[{p^{2}\over 2m_{i}}+\varepsilon_{i}\right]a^{\dagger}_{i\bf p}a_{i\bf p}, \quad i=1,2.
\end{equation}
The interaction Hamiltonian, as previously, is given by $V=V_{bb}+V_{bf}+V_{ff}$, where
\begin{equation}\label{eq:2.3}
V_{bb}={1\over 2\mathcal{V}}\sum_{{\bf p}_{1}...{\bf p}_{4}}g({\bf p}_{14})\,b^{\dagger}_{{\bf p}_{1}}b^{\dagger}_{\bf p_{2}}b_{\bf p_{3}}b_{{\bf p}_{4}}\,\delta_{{\bf p}_{1}+{\bf p}_{2},{\bf p}_{3}+{\bf p}_{4}},
\end{equation}
with the interaction amplitude
\begin{eqnarray}
g({\bf p}_{14})=\sigma^{(2)}({\bf p}_{14})\nu_{11}({\bf p}_{14})\sigma^{*(2)}({\bf p}_{14}) +\sigma^{*(1)}({\bf p}_{14})\nu_{22}({\bf p}_{14})\sigma^{(1)}({\bf p}_{14}) \nonumber  \\
+\sigma^{(2)}({\bf p}_{14})\nu_{12}({\bf p}_{14})\sigma^{(1)}({\bf p}_{14})+\sigma^{*(1)}({\bf p}_{14})\nu_{21}({\bf p}_{14})\sigma^{*(2)}({\bf p}_{14}) \label{eq:2.4}
\end{eqnarray}
and
\begin{equation}\label{eq:2.5}
\sigma^{(i)}({\bf p})=\int d{\bf x}\,\phi^{*}({\bf x})\phi({\bf x})\exp\left[ i{{\bf p}{\bf x}}{m_{i}\over \hbar M}\right].
\end{equation}
The operator $V_{bf}$ (see Eq. (\ref{eq:1.14})) describing the interaction of bound states with fermionic atoms is written as follows:
\begin{eqnarray}
V_{bf}&={1\over \mathcal{V}}\sum_{{\bf p}_{1}...{\bf p}_{4}}v({\bf p}_{21})\,b^{\dagger}_{{\bf p}_{1}}b_{\bf p_{2}}a^{\dagger}_{1\bf p_{3}}a_{1{\bf p}_{4}}\,\delta_{{\bf p}_{2}-{\bf p}_{1},{\bf p}_{3}-{\bf p}_{4}} \nonumber \\
&+{1\over \mathcal{V}}\sum_{{\bf p}_{1}...{\bf p}_{4}}u({\bf p}_{21})\,b^{\dagger}_{{\bf p}_{1}}b_{\bf p_{2}}a^{\dagger}_{2\bf p_{3}}a_{2{\bf p}_{4}}\,\delta_{{\bf p}_{2}-{\bf p}_{1},{\bf p}_{3}-{\bf p}_{4}}, \label{eq:2.6}
\end{eqnarray}
where
\begin{eqnarray}
v({\bf p}_{21})=\nu_{11}({\bf p}_{21})\sigma^{*(2)}({\bf p}_{21})+\nu_{21}({\bf p}_{21})\sigma^{(1)}({\bf p}_{21}), \nonumber \\
u({\bf p}_{21})=\nu_{22}({\bf p}_{21})\sigma^{(1)}({\bf p}_{21})+\nu_{12}({\bf p}_{21})\sigma^{*(2)}({\bf p}_{21}). \label{eq:2.7}
\end{eqnarray}
The operator $V_{ff}$, as before, is given by Eq. (\ref{eq:1.16}).

We now define the total particle number operator. Since each diatomic bound state contains only two interspecies fermionic atoms, the total particle number operator is written in the form,
\begin{equation} \label{eq:2.8}
N=N_{1}+N_{2}, \quad N_{i}=\sum_{\bf p}a^{\dagger}_{i\bf p}a_{i\bf p}+\sum_{\bf p}b^{\dagger}_{\bf p}b_{\bf p}, \quad i=1,2.
\end{equation}
Note that the above Hamiltonian considers the bound states as stable and does not take into account the processes associated with conversion of atoms into a bound state and its reconversion to unbound fermionic atoms, so that we have $[H,N_{i}]=0$. Therefore, the total number of atoms of both components is conserved. Since the operator of the total bound state number $N_{b}=\sum_{{\bf p}\neq 0}b_{\bf p}^{\dagger}b_{\bf p}$ commutes with the Hamiltonian, it would seem, from the mathematical point of view, that $N_{b}$ should be treated on an equal footing with $N_{i}$. However, it is worth stressing that only $N_{i}$ can be specified arbitrary in the state of statistical equilibrium. Due to this fact, all thermodynamic characteristics of the system, including the numbers of bound states and unbound fermionic atoms, should be expressed in terms of $N_{i}$ and temperature $T$. Hence, in spite of the fact that $[N_{b},H]=0$, the total bound state number is not conserved when thermodynamic characteristics of the system (e.g., temperature) are varied. From the physical point of view, this is a difference between the total number of fermionic atoms of both species and total numbers of bound states as well as unbound fermions. Such a situation is typical in the theory of chemical reactions, when a system consists of two reagents and one product of reaction (see, e.g., \cite{Landau2}). Therefore, regardless of the character of interaction, in order to describe the system in the grand canonical ensemble, one needs to introduce only two Lagrange multipliers or two chemical potentials $\mu_{1}$ and $\mu_{2}$ associated with conserved quantities $N_{1}$ and $N_{2}$, so that
\begin{equation}\label{eq:2.9}
\mu_{1}N_{1}+\mu_{2}N_{2}=\mu\sum_{\bf p}b^{\dagger}_{\bf p}b_{\bf p}+\mu_{1}\sum_{\bf p}a^{\dagger}_{1\bf p}a_{1\bf p}+\mu_{2}\sum_{\bf p}a^{\dagger}_{2\bf p}a_{2\bf p}, \quad \mu=\mu_{1}+\mu_{2}.
\end{equation}
As we see, the chemical potential $\mu$ associated with bound states is expressed in terms of the chemical potentials of independent fermionic components, which is a consequence of the so-called Gibbs rule. For this reason, the system under consideration differs from the ordinary three-component system whose description requires introducing three chemical potentials.

Suppose the temperature of the system is so low that the bound states are Bose condensed.
This means that their number with zero momentum is a macroscopic value, i.e. it is proportional to the volume of the system, $N_{0}=\langle b_{0}^{\dagger}b_{0}\rangle\sim\mathcal{V}$, where $\langle...\rangle$ denotes the averaging with equilibrium statistical operator. If one introduces the operators $\beta_{0}=b_{0}/\sqrt{\mathcal{V}}$ and $\beta_{0}^{\dagger}=b_{0}^{\dagger}/\sqrt{\mathcal{V}}$ so that $\langle\beta_{0}^{\dagger}\beta_{0}\rangle\sim 1$, they satisfy the commutation relations $[\beta_{0}, \beta_{0}^{\dagger}]=1/\mathcal{V}$ and $[\beta_{0}, b_{{\bf p}\neq 0}^{\dagger}]=0$. Therefore, as $\mathcal{V}\to\infty$, the quantities $\beta_{0}$, $\beta_{0}^{\dagger}$, and, consequently, $b_{0}$, $b_{0}^{\dagger}$ behave as $c$ -- numbers, which are different from zero. Such a treatment of the condensate particles was presented by Bogoliubov when constructing the special perturbative approach for a weakly interacting Bose gas with condensate \cite{Bogoliubov}. More rigorous justification of this treatment can be given using the Bogoliubov quasi-average concept and the principle of spatial correlation weakening \cite{Akh-Pel}. Therefore, $b^{\dagger}_{0}$ and $b_{0}$ are replaced by $N_{0}^{1/2}$ in all operators of relevant physical quantities, where $N_{0}$ is the number of diatomic bound states with momentum ${\bf p}=0$. This replacement results in the following expression for the kinetic energy operator (see Eq. (\ref{eq:2.2})):
\begin{equation} \label{eq:2.10}
H_{0}(N_{0})=\varepsilon N_{0}+\sum_{{\bf p}\neq 0}\left[{p^{2}\over 2M}+\varepsilon\right]b^{\dagger}_{{\bf p}}b_{{\bf p}}+\sum_{i,{\bf p}}\left[{p^{2}\over 2m_{i}}+\varepsilon_{i}\right]a^{\dagger}_{i{\bf p}}a_{i{\bf p}}.
\end{equation}
In a similar manner, the interaction Hamiltonian $V=V_{bb}+V_{bf}+V_{ff}$ is reduced to
\begin{equation} \label{eq:2.11}
V(N_{0})=f(N_{0})+{\partial f(N_{0})\over\partial N_{0}}N'_{b}+N_{0}V^{(2)}+N_{0}^{1/2}V^{(3)}+V^{(4)},
\end{equation}
where
\begin{equation} \label{eq:2.12}
f(N_{0})={1\over 2\mathcal{V}}g(0)N_{0}^{2}, \quad N'_{b}=\sum_{{\bf p}\neq 0}b^{\dagger}_{{\bf p}}b_{{\bf p}}.
\end{equation}
The operator $V^{(2)}$, quadratic in creation and annihilation operators of bosons and fermions, has the form
\begin{eqnarray}
V^{(2)}&={1\over \mathcal{V}}\sum_{{\bf p}\neq 0}g({\bf p})b^{\dagger}_{{\bf p}}b_{{\bf p}}+{1\over 2\mathcal{V}}\sum_{{\bf p}\neq 0}g({\bf p})\left[b^{\dagger}_{{\bf p}}b^{\dagger}_{-{\bf p}}+b_{{\bf p}}b_{-{\bf p}}\right] \nonumber
\\
&+{1\over \mathcal{V}}\sum_{\bf p}v(0)a^{\dagger}_{1{\bf p}}a_{1{\bf p}}+{1\over \mathcal{V}}\sum_{\bf p}u(0)a^{\dagger}_{2{\bf p}}a_{2{\bf p}}. \label{eq:2.13}
\end{eqnarray}
The operators $V^{(3)}$ and $V^{(4)}$ are given by
\begin{eqnarray}
\fl V^{(3)}={1\over \mathcal{V}}\sum_{{{\bf p}_{1}\neq 0},{\bf p}_{2},{\bf p}_{3}}\left[v({\bf p}_{1})\,b_{{\bf p}_{1}}a^{\dagger}_{1{\bf p}_{2}}a_{1{\bf p}_{3}}\delta_{{\bf p}_{1},{\bf p}_{2}-{\bf p}_{3}}+v(-{\bf p}_{1})\,b^{\dagger}_{{\bf p}_{1}}a^{\dagger}_{1{\bf p}_{2}}a_{1{\bf p}_{3}}\delta_{-{\bf p}_{1},{\bf p}_{2}-{\bf p}_{3}}\right] \nonumber \\
+{1\over \mathcal{V}}\sum_{{{\bf p}_{1}\neq 0},{\bf p}_{2},{\bf p}_{3}}\left[u({\bf p}_{1})\,b_{{\bf p}_{1}}a^{\dagger}_{2{\bf p}_{2}}a_{2{\bf
p}_{3}}\delta_{{\bf p}_{1},{\bf p}_{2}-{\bf p}_{3}}+u(-{\bf p}_{1})\,b^{\dagger}_{{\bf p}_{1}}a^{\dagger}_{2{\bf p}_{2}}a_{2{\bf
p}_{3}}\delta_{-{\bf p}_{1},{\bf p}_{2}-{\bf p}_{3}}\right] \nonumber \\
+{1\over \mathcal{V}}\sum_{{\bf p}_{1}...{\bf p}_{3}\neq 0}g({\bf p}_{13})\left[b^{\dagger}_{{\bf p}_{1}}b_{{\bf p}_{2}}b_{{\bf p}_{3}}\delta_{{\bf p}_{1},{\bf p}_{2}+{\bf p}_{3}}+b^{\dagger}_{{\bf p}_{1}}b^{\dagger}_{{\bf p}_{2}}b_{{\bf p}_{3}}\delta_{{\bf p}_{1}+{\bf p}_{2},{\bf p}_{3}}\right]   \label{eq:2.14}
\end{eqnarray}
and
\begin{eqnarray}
V^{(4)}&={1\over 2\mathcal{V}}\sum_{{\bf p}_{1}...{\bf p}_{4}\neq 0}g({\bf p}_{14})\,b^{\dagger}_{{\bf p}_{1}}b^{\dagger}_{{\bf p}_{2}}b_{{\bf p}_{3}}b_{{\bf p}_{4}}\,\delta_{{\bf p}_{1}+{\bf p}_{2},{\bf p}_{3}+{\bf p}_{4}} \nonumber \\
&+{1\over \mathcal{V}}\sum_{{\bf p}_{1},{\bf p}_{2}\neq 0,{\bf p}_{3},{\bf p}_{4}}b^{\dagger}_{{\bf p}_{1}}b_{{\bf p}_{2}}\left[a^{\dagger}_{1{\bf p}_{3}}a_{1{\bf p}_{4}}v({\bf p}_{21})+a^{\dagger}_{2{\bf p}_{3}}a_{2{\bf p}_{4}}u({\bf p}_{21})\right]\delta_{{\bf p}_{2}-{\bf p}_{1},{\bf p}_{3}-{\bf p}_{4}} \nonumber \\
&+{1\over 2\mathcal{V}}\sum_{{\bf p}_{1}...{\bf p}_{4}}a_{1{\bf p}_{1}}^{\dagger}a_{1{\bf p}_{2}}^{\dagger}a_{1{\bf p}_{3}}a_{1{\bf p}_{4}}\nu_{11}({\bf p}_{14})\,\delta_{{\bf p}_{1}+{\bf p}_{2},{\bf p}_{3}+{\bf p}_{4}} \nonumber \\
&+{1\over 2\mathcal{V}}\sum_{{\bf p}_{1}...{\bf p}_{4}}a_{2{\bf p}_{1}}^{\dagger}a_{2{\bf p}_{2}}^{\dagger}a_{2{\bf p}_{3}}a_{2{\bf p}_{4}}\nu_{22}({\bf p}_{14})\,\delta_{{\bf p}_{1}+{\bf p}_{2},{\bf p}_{3}+{\bf p}_{4}} \nonumber \\
&+{1\over\mathcal{V}}\sum_{{\bf p}_{1}...{\bf p}_{4}}a_{1{\bf p}_{1}}^{\dagger}a_{1{\bf p}_{2}}a_{2{\bf p}_{3}}^{\dagger}a_{2{\bf p}_{4}}\nu_{12}({\bf p}_{12})\,\delta_{{\bf p}_{1}-{\bf p}_{2},{\bf p}_{4}-{\bf p}_{3}}. \label{eq:2.15}
\end{eqnarray}
The same replacement, $b_{0},\,b_{0}^{\dagger}\to N_{0}^{1/2}$, should also be performed in Eq. (\ref{eq:2.9}). Then, the Gibbs statistical operator corresponding to the grand canonical ensemble takes the form
\begin{equation} \label{eq:2.16}
w(N_{0})=\exp\left[\Omega-\beta(H(N_{0})-\mu N_{0}-\mu N'_{b}-\mu_{1}N_{1}'-\mu_{2}N_{2}')\right],
\end{equation}
where $H(N_{0})=H_{0}(N_{0})+V(N_{0})$ and $N'_{i}$ is the number of unbound fermionic atoms of the first or second species,
\begin{equation} \label{eq:2.17}
N'_{i}=\sum_{\bf p}a^{\dagger}_{i\bf p}a_{i\bf p}, \quad i=1,2.
\end{equation}
The grand thermodynamic potential $\Omega$ as a function of reciprocal temperature $\beta=1/T$, chemical potentials $\mu_{1}$ and $\mu_{2}$, and number of condensed bound states $N_{0}$ is found from the normalization condition ${\rm Sp}\,w(N_{0})=1$, where the trace is taken in the space of occupation numbers of bosons with momentum ${\bf p}\neq 0$ and fermions with all possible values of momentum. Taking into account Eq. (\ref{eq:2.16}) and normalization condition, one obtains
$$
{\partial\Omega\over\partial N_{0}}=-\beta\left[\mu-\varepsilon-{\rm Sp}w(N_{0}){\partial V(N_{0})\over\partial N_{0}}\right], \quad \mu=\mu_{1}+\mu_{2}.
$$
Using the Bogoliubov method of quasi-averages \cite{Bog,Bogol-Pel,Akh-Pel}, which was elaborated for describing the systems with spontaneously broken symmetries, one can prove, in a mathematically rigorous way, that the expression in square brackets vanishes \cite{Akh-Pel}, so that $N_{0}$ is found from the minimum condition for thermodynamic potential,
\begin{equation}\label{eq:2.18}
{\rm Sp}\,w(N_{0}){\partial V(N_{0})\over\partial N_{0}}+\varepsilon-\mu=0.
\end{equation}
Here the chemical potential $\mu$ should be expressed in terms of $N_{1}$ and $N_{2}$ and, consequently, the number of condensed bound states is determined by the numbers of fermionic atoms of both species and temperature, $N_{0}=N_{0}(N_{1},N_{2},T)$.

It is worth stressing that all the found relationships, including Eq. (\ref{eq:2.18}), are exact. When obtaining them, we have only used the replacement of corresponding operators by $c$ -- numbers and have not developed any perturbative approach. Recall that, along with low temperatures, the interaction between all the components of the system should be weak enough to ensure the stability of the bound states in equilibrium (see estimates in section 4).
Then $N_{0}/\mathcal{V}$ is a large parameter, since when $g({\bf p})\to 0$ and $T\to 0$, all the bound states are Bose condensed. Thus, in Eq. (\ref{eq:2.11}), $f(N_{0})$ is the largest term and the next ones in order of magnitude are $[{\partial f/\partial N_{0}}]N'_{b}$ and $N_{0}V^{(2)}$. We neglect two other terms $N_{0}^{1/2}V^{(3)}$ and $V^{(4)}$, in Eq. (\ref{eq:2.11}), which must be taken into account when describing the interaction effects between quasi-particles. The role of these terms within the developed perturbative approach is discussed in the section 4. Therefore, replacing $V(N_{0})$ with $f(N_{0})$, one obtains from Eq. (\ref{eq:2.18}) the following equation:
\begin{equation}\label{eq:2.19}
\mu\approx\varepsilon+{\partial f(N_{0})\over\partial N_{0}}=\varepsilon+{g(0)N_{0}\over \mathcal{V}}, \quad \mu=\mu_{1}+\mu_{2}.
\end{equation}
This provides a relation between $\mu$ and $N_{0}$ in the leading approximation. Next, retaining the terms up to $N_{0}V^{(2)}$ in Eq. (\ref{eq:2.11}), and eliminating $\mu$ given by Eq. (\ref{eq:2.19}), one finds the following Gibbs statistical operator (see Eq. (\ref{eq:2.16})):
\begin{equation} \label{eq:2.20}
w(N_{0})\approx w^{(2)}(N_{0})=\exp\left[\Omega_{0}-\beta\mathcal{H}^{(2)}(N_{0})\right],
\end{equation}
where
\begin{eqnarray}
\fl \mathcal{H}^{(2)}(N_{0})=-f(N_{0})+\sum_{{\bf p}\neq 0}\left[{p^{2}\over 2M}+{g({\bf p})N_{0}\over \mathcal{V}}\right]b^{\dagger}_{\bf p}b_{\bf p}+{N_{0}\over 2\mathcal{V}}\sum_{{\bf p}\neq 0}g({\bf p})\left[b^{\dagger}_{\bf p}b^{\dagger}_{-{\bf p}}+b_{\bf p}b_{-{\bf p}}\right] \nonumber \\
+\sum_{i,\bf p}\left[{p^{2}\over 2m_{i}}-\tilde{\mu}_{i}\right]a^{\dagger}_{i{\bf p}}a_{i{\bf p}}, \quad i=1,2,
\end{eqnarray}
and
\begin{equation} \label{eq:2.22}
\tilde{\mu}_{1}=\mu_{1}-\varepsilon_{1}-{v(0)N_{0}\over\mathcal{V}}, \quad \tilde{\mu}_{2}=\mu_{2}-\varepsilon_{2}-{u(0)N_{0}\over\mathcal{V}}.
\end{equation}
The grand thermodynamic potential $\Omega_{0}$ in Eq. (\ref{eq:2.20}) is found from the normalization condition ${\rm Sp}\,w_{0}(N_{0})=1$. Note that $\Omega$ in Eq. (\ref{eq:2.16}) coincides with $\Omega_{0}$ in the approximation under consideration. The operator $\mathcal{H}^{(2)}(N_{0})$ consists of bosonic and fermionic parts with respect to creation and annihilation operators. The fermionic part has a diagonal form, whereas the bosonic part should be diagonalized.

To this end, we introduce the unitary transformation $U$ ($UU^{\dagger}=1$) that acts on bosonic operators and does not affect the fermionic operators. This transformation reduces $\mathcal{H}^{(2)}(N_{0})$ to the following diagonal form:
\begin{equation}\label{eq:2.23}
U\mathcal{H}^{(2)}(N_{0})U^{\dagger}=-f(N_{0})+\sum_{{\bf p}\neq 0}\omega_{\bf p}b^{\dagger}_{\bf p}b_{\bf p}+\sum_{i,\bf p}\left[{p^{2}\over 2m_{i}}-\tilde{\mu}_{i}\right]a^{\dagger}_{i{\bf p}}a_{i{\bf p}},
\end{equation}
where $\omega_{\bf p}$ is the single-particle excitation spectrum. For the diagonalization of $\mathcal{H}^{(2)}(N_{0})$, it is sufficient to restrict ourselves to unitary operators $U$, which mix up the operators $b^{\dagger}_{-{\bf p}}$ and $b_{\bf p}$:
\begin{equation}\label{eq:2.24}
Ub^{\dagger}_{\bf p}U^{\dagger}=\mathcal{U}_{\bf p}b^{\dagger}_{\bf p}+\mathcal{V}_{\bf p}b_{-{\bf p}}, \quad Ub_{\bf p}U^{\dagger}=\mathcal{U}_{\bf p}b_{\bf p}+\mathcal{V}_{\bf p}b^{\dagger}_{-{\bf p}}.
\end{equation}
These new operators must satisfy the bosonic commutation relation. This requirement results in the following well-known relationships for $\mathcal{U}_{\bf p}$ and $\mathcal{V}_{\bf p}$:
\begin{equation}\label{eq:2.25}
\mathcal{U}_{\bf p}^{2}-\mathcal{V}_{\bf p}^{2}=1, \quad \mathcal{U}_{\bf p}\mathcal{V}_{-{\bf p}}-\mathcal{V}_{\bf p}\mathcal{U}_{-{\bf p}}=0.
\end{equation}
Next, performing the standard diagonalization procedure \cite{Bogoliubov}, one finds the quasi-particle spectrum,
\begin{equation}\label{eq:2.26}
\omega_{\bf p}=\bigg[\left({p^{2}\over 2M}\right)^{2}+{p^{2}\over M}g({\bf p})n_{0}\bigg]^{1/2},
\end{equation}
as well as the functions $\mathcal{U}_{\bf p}$ and $\mathcal{V}_{\bf p}$, which define the unitary transformation $U$,
\begin{eqnarray}
\mathcal{U}_{\bf p}^{2}={[p^{2}/2M+g({\bf p})n_{0}+\omega_{\bf p}]^{2}\over{[p^{2}/2M+g({\bf p})n_{0}+\omega_{\bf p}]^{2}-g^{2}({\bf p})n_{0}^{2}}}, \nonumber \\
\mathcal{V}_{{\bf p}}^{2}={g^{2}({\bf p})n_{0}^{2}\over{2\omega_{\bf p}[{p^{2}/2M}+g({\bf p })n_{0}+\omega_{\bf p}]}}\label{eq:2.27},
\end{eqnarray}
where $n_{0}=N_{0}/\mathcal{V}$ is the density of the condensed bound states.
The form of the spectrum given by Eq. (\ref{eq:2.26}) coincides with that obtained by Bogoliubov within the model for a weakly interacting Bose gas \cite{Bogoliubov}. However, it has some specific features. It particular, according to Eq. (\ref{eq:2.4}), the interaction amplitude $g({\bf p})$ is expressed through the interactions of unbound fermionic atoms and wave functions of the bound states. Moreover, the condensate density $n_{0}$, given by Eq. (\ref{eq:2.19}), depends on the densities of fermionic atoms of both species. Note that if the Hamiltonian contains the terms describing the conversion of two atoms into a bound state and its reconversion to unbound atoms, then the single-particle spectrum may exhibit a gap, at least for a pure bosonic system \cite{Annals,Pel-Pel-Pol}.

Thus, the unitary transformation given by Eqs. (\ref{eq:2.24}) reduces the Gibbs statistical operator (\ref{eq:2.20}) to the diagonal form,
\begin{equation} \label{eq:2.28}
Uw^{(2)}(N_{0})U^{\dagger}=\exp\left\{\tilde{\Omega}_{0}-\beta\sum_{{\bf p}\neq 0}\omega_{\bf p}b^{\dagger}_{{\bf p}}b_{{\bf
p}}-\beta\sum_{i,{\bf p}}\left[{p^{2}\over 2m_{i}}-\tilde{\mu}_{i}\right]a^{\dagger}_{i{\bf p}}a_{i{\bf p}}\right\},
\end{equation}
where $\tilde{\Omega}_{0}=\Omega_{0}-\beta f(N_{0})$. This statistical operator allows one to find the average values of physical quantities. The energy of the system corresponding to the quadratic approximation of Hamiltonian is
\begin{equation}\label{eq:2.29}
E={\rm Sp}\,w^{(2)}(N_{0})H^{(2)}(N_{0})={\rm Sp}\,Uw^{(2)}(N_{0})U^{\dagger}UH^{(2)}(N_{0})U^{\dagger},
\end{equation}
where $H^{(2)}(N_{0})$ is a true Hamiltonian related to $\mathcal{H}^{(2)}(N_{0})$ by
\begin{equation}\label{eq:2.30}
H^{(2)}(N_{0})=\mathcal{H}^{(2)}(N_{0})\big\vert_{\mu_{1}=\mu_{2}=0}+{\varepsilon}N_{0}+{g(0)N_{0}^{2}\over \ \mathcal{V}}.
\end{equation}
It is diagonalized by the same unitary transformation and its bosonic part coincides with the Hamiltonian obtained in \cite{Bogoliubov}. The trace in Eq. (\ref{eq:2.29}) can be easily computed. In the next section we provide the corresponding result for the ground state energy, when the trace is computed at zero temperature.

Let us find the explicit expressions for the total number of fermionic atoms of both species. To this end, we address Eq. (\ref{eq:2.8}), which gives
$$
N_{i}=N_{0}+\sum_{{\bf p}\neq 0}{\rm Sp}\,w^{(2)}(N_{0})\,b^{\dagger}_{{\bf p}}b_{{\bf p}}+\sum_{{\bf p}}{\rm Sp}\,w^{(2)}(N_{0})\,a^{\dagger}_{i{\bf p}}a_{i{\bf p}}.
$$
Performing here the unitary transformation under the spur sign and using Eqs. (\ref{eq:2.24}), (\ref{eq:2.25}), (\ref{eq:2.28}), one obtains
\begin{equation}\label{eq:2.31}
N_{i}=N_{0}+\sum_{{\bf p}\neq 0}\left[(2\mathcal{U}_{\bf p}^{2}-1)f_{b\bf p}+\mathcal{U}_{\bf p}^{2}-1\right]+\sum_{{\bf p}}f_{i\bf p},
\end{equation}
where
\begin{equation}\label{eq:2.31'}
f_{b\bf p}={1\over{e^{\beta\omega_{\bf p}}-1}}, \qquad f_{i\bf p}={1\over {e^{\beta\left[(p^{2}/2m_{i})-\tilde{\mu}_{i}\right]}+1}}.
\end{equation}
In thermodynamic equilibrium, Eqs. (\ref{eq:2.31}), (\ref{eq:2.31'}) provide the relationship between the numbers of bound states and unbound fermionic atoms if the temperature $T$ and total number of fermions $N_{i}$ are given. If one changes the temperature, after a relaxation time, the system comes to another equilibrium state with other numbers of unbound fermions and bound states but with the same values of $N_{i}$. In this sense, as discussed above in this section, the number of bound states (and also unbound fermions) is not a conserved quantity at any temperature in spite of the fact that the corresponding operators commute with the Hamiltonian. Using now Eq. (\ref{eq:2.27}) to eliminate $\mathcal{U}_{\bf p}$ and replacing summation with integration, we have
\begin{equation} \label{eq:2.32}
N_{i}=N_{0}+{\mathcal{V}\over 2\pi^{2}\hbar^{3}}(I_{i1}+I_{2}+I_{3}), \quad i=1,2,
\end{equation}
where
\begin{equation} \label{eq:2.33}
I_{i1}=\int_{0}^{\infty}dp\,p^{2}{1\over {e^{\beta(p^{2}/2m_{i}-\tilde{\mu}_{i})}+1}},
\end{equation}
\begin{equation} \label{eq:2.34}
I_{2}={1\over 2}\int_{0}^{\infty}dp\,p^{2}{g^{2}({\bf p})n_{0}^{2}\over{\omega_{\bf p}[{p^{2}/2M}+g({\bf p })n_{0}+\omega_{\bf p}]}},
\end{equation}
\begin{equation}\label{eq:2.35}
I_{3}={1\over 2}\int_{0}^{\infty}dp\,p^{2}\,{1\over{e^{\beta\omega_{\bf p}}-1}}{{[{p^{2}/2M}+g({\bf p })n_{0}+\omega_{\bf p}]^{2}+g({\bf p })n_{0}}\over{\omega_{\bf p}[{p^{2}/2M}+g({\bf p })n_{0}+\omega_{\bf p}]}}.
\end{equation}

So far we have not specified the form of the interaction between the atoms and their bound states. In this sense, the obtained results are general. However, to perform concrete calculations we should specify the interaction between atoms. For ultracold dilute gases, the simplest way is to consider the elastic scattering of slow atoms, when their interactions are characterized by the corresponding scattering lengths \cite{Pethick,Pitaevskii}. We assume that such treatment of interactions is acceptable for our problem, i.e. when ${\bf p}\to 0$, all the interaction amplitudes $\nu_{ij}({\bf p})$ tend to their constant values, which are expressed through the scattering lengths,
\begin{equation}\label{eq:2.36}
\nu_{11}={4\pi\hbar^{2}\over m_{1}}a_{11}, \quad \nu_{22}={4\pi\hbar^{2}\over m_{2}}a_{22}, \quad \nu_{12}={2\pi\hbar^{2}\over m_{*}}a_{12},
\end{equation}
where $m_{*}=m_{1}m_{2}/(m_{1}+m_{2})$ is the reduced mass. Moreover, the amplitude $g({\bf p})$, given by Eq. (\ref{eq:2.4}), does not depend on momentum, since $\sigma({\bf p})\to 1$ when ${\bf p}\to 0$ (see Eq. (\ref{eq:2.5})). It is expressed through the scattering lengths $a_{ij}$ ($a_{12}=a_{21}$),
\begin{equation}\label{eq:2.37}
g=4\pi\hbar^{2}\left[{a_{11}\over m_{1}}+{a_{22}\over m_{2}}+{{(m_{1}+m_{2})}\over {m_{1}m_{2}}}a_{12}\right].
\end{equation}
In a similar manner, Eqs. (\ref{eq:2.7}) give
\begin{eqnarray}
v=4\pi\hbar^{2}\left[{a_{11}\over m_{1}}+{{(m_{1}+m_{2})}\over {m_{1}m_{2}}}{a_{21}\over 2}\right], \nonumber \\
u=4\pi\hbar^{2}\left[{a_{22}\over m_{2}}+{{(m_{1}+m_{2})}\over {m_{1}m_{2}}}{a_{12}\over 2}\right], \label{eq:2.38}
\end{eqnarray}
and $u+v=g$. In the subsequent analysis, we assume that all interactions are characterized by the scattering lengths. It is worth stressing that the replacement of real interaction potentials by their constant values (or by scattering lengths) is a not so "inoffensive" approximation. As for the generality of a theory, we lose the information regarding the microscopic characteristics of the system: the effect of the bound state wave function on the character of interaction between heteronuclear 'molecules' is neglected (compare Eqs. (\ref{eq:2.4}), (\ref{eq:2.5}) with (\ref{eq:2.37})). At the same time, Eqs. (\ref{eq:2.4}), (\ref{eq:2.5}) show that this effect can be significant. In addition, the above-mentioned approximation results in divergences of integrals in the ground state energy so that it is necessary to use the renormalization of the coupling constant (see, e.g., \cite{Pethick,Landau} and next section). However, to estimate the losses associated with the approximation under consideration, one needs to solve the declared problem with more realistic interaction potentials between fermions and then to compare the results with those obtained in this work. The realization of such a procedure represents a quite complicated separate problem which, can be solved employing complex numerical methods.

Note that the low-energy collisions of fermionic atoms can be described by $s$ -- wave scattering length only if they scatter in different internal (spin) states. However, in experiments with heteronuclear molecules, each of the two fermionic components is polarized in some internal state. For example, as was already mentioned, when creating ${}^{6}{\rm Li}$--${}^{40}{\rm K}$ molecules \cite{Voigt}, the fermionic atoms are prepared in the states K$|F=9/2,m_{F}=-5/2\rangle$ and Li$|F=1/2,m_{F}=1/2\rangle$, where $F$ and $m_{F}$ are the total spin and its projection, respectively. The internal energies $\varepsilon_{1}$ and $\varepsilon_{2}$ introduced above correspond to such states. Therefore, due to Fermi statistics, two atoms from the same component must scatter with odd values of the angular momentum. Hence, their collisions should be described, at least, by $p$ -- wave scattering, whose contribution to the scattering amplitude is small for low-energy atoms. For this reason, below we neglect the quantities $\nu_{11}$ and $\nu_{22}$, while the interspecies atomic interaction is described by $s$ -- wave scattering length $a_{12}$. Thus, Eqs. (\ref{eq:2.37}), (\ref{eq:2.38}) give
\begin{equation} \label{eq:2.37'}
g\equiv {4\pi\hbar^{2}\over M}a_{b}, \quad a_{b}={(m_{1}+m_{2})^{2}\over m_{1}m_{2}}a_{12},
\end{equation}
è
\begin{equation}\label{eq:2.37''}
u=v={g\over 2},
\end{equation}
where $a_{b}$ -- is the scattering length of the bound states. For $m_{1}\ll m_{2}$, one finds
\begin{equation}\label{eq:2.37'''}
a_{b}\approx \left(2+{m_{2}\over m_{1}}\right)a_{12}, \quad m_{1}\ll m_{2}.
\end{equation}
Equation (\ref{eq:2.37'''}) yields that $a_{b}\gg a_{12}$. This fact allows one to conclude that the range of the bound states is much smaller than their scattering length. Indeed, even if one considers that the range of the bound states is of the order of the range of potential characterized by the scattering length $a_{12}$, one can easily find that $a_{b}$ is much larger than the range of the bound state. Below we analyze the basic equations using the data for a degenerate mixture of ${}^{6}{\rm Li}$ and ${}^{173}{\rm Yb}$ atoms \cite{Hara}. In this case $m_{1}\ll m_{2}$ and Eq. (\ref{eq:2.37'''}) is valid. Note that even for ${}^{6}{\rm Li}$--${}^{40}{\rm K}$ molecules \cite{Voigt}, Eq. (\ref{eq:2.37'''}) can be considered to be true since $m_{2}/m_{1}\approx 6.7$.

We now focus on integrals entering Eq. (\ref{eq:2.32}) for the total number of fermionic atoms of both species. The integral given by Eq. (\ref{eq:2.34}) is independent of temperature and can be computed exactly. To calculate the integral given by Eq. (\ref{eq:2.33}), let us consider the temperatures at which both components of fermionic atoms are degenerate, i.e. when the conditions ${\tilde{\mu}_{i}/T}\gg 1$ are satisfied. In this case we can apply the low-temperature expansion, which is used, e.g., when calculating the heat capacity of degenerate electron gas \cite{Landau2}. Finally, the integral given by Eq. (\ref{eq:2.35}) can be easily computed when $gn_{0}/T\gg 1$. Therefore, the result of computations reads
\begin{equation} \label{eq:2.39}
I_{i1}\approx {1\over 3}(2m_{i}\tilde{\mu}_{i})^{3/2}+{\pi^{2}\over 6\sqrt{2}}{m_{i}^{3/2}\sqrt{\tilde{\mu}_{i}}}T^{2}, \quad {\tilde{\mu}_{i}\over T}\gg 1,
\end{equation}
\begin{equation} \label{eq:2.40}
I_{2}={2\over 3}(Mgn_{0})^{3/2},
\end{equation}
\begin{equation} \label{eq:2.41}
I_{3}\approx{\pi^{2}\over 6}{M^{3/2}\over\sqrt{gn_{0}}}T^{2}, \quad {gn_{0}\over T}\gg 1.
\end{equation}
The computed integrals, along with Eq. (\ref{eq:2.32}), allow one to express the chemical potentials $\mu_{1}$ and $\mu_{2}$ through $N_{1}$ and $N_{2}$, respectively.

\section{Zero temperature}

An approximate second quantization method, which is based on replacement of creation and annihilation operators with zero momentum by $c$ -- numbers, implies the condensate density of bound states is different from zero. In particular, in the original Bogoliubov theory \cite{Bogoliubov}, the number of condensed atoms is close to their total number. However, this original theory cannot predict the lower and upper values of the condensate density. Because of the presence of fermionic atoms in the system, the proposed theory makes it possible to carry out such estimates, at least at zero temperature. So far, in Eq. (\ref{eq:2.11}), we have taken into account the terms up to $N_{0}V^{(2)}$ inclusive, and neglected $N_{0}^{1/2}V^{(3)}$ and $V^{(4)}$, which describe the interaction between quasi-particles. The ground state energy corresponding to the quadratic Hamiltonian can be computed from Eq. (\ref{eq:2.29}). However, to estimate the variation range of $n_{0}$ at fixed densities of fermionic atoms, we take into account the term $V^{(4)}$ given by Eq. (\ref{eq:2.15}). It is evident that ${\rm Sp}\,w^{(2)}(N_{0})V^{(3)}=0$. Thus, the ground state energy is written as
\begin{equation} \label{eq:3.1}
\mathcal{E}_{0}=\mathcal{E}_{0}^{(0)}+\mathcal{E}_{0}^{(2)}+\mathcal{E}_{0}^{(4)}.
\end{equation}
Here $\mathcal{E}_{0}^{(0)}$ is a $c$ -- number part of the Hamiltonian,
\begin{equation} \label{eq:3.2}
\mathcal{E}_{0}^{(0)}=-|\varepsilon| n_{0}+{gn_{0}^{2}\over 2}.
\end{equation}
Two other terms represent the computed traces (at zero temperature) for the quadratic part of the Hamiltonian given by Eq. (\ref{eq:2.30}) and for $V^{(4)}$, respectively. Taking into account Eq. (\ref{eq:2.37''}) these terms are found to be
\begin{eqnarray}
\mathcal{E}_{0}^{(2)}&=-{1\over 2\mathcal V}\sum_{{\bf p}\neq 0}\left[{p^{2}\over 2M}+gn_{0}-\omega_{\bf p}\right]+{3\over 10}(6\pi^{2})^{2/3}{\hbar^{2}\over m_{1}}{n'_{1}}^{5/3} \nonumber \\ &-\left(|\varepsilon_{1}|-{g\over 2}n_{0}\right)n'_{1}+{3\over 10}(6\pi^{2})^{2/3}{\hbar^{2}\over m_{2}}{n'_{2}}^{5/3}-\left(|\varepsilon_{2}|-{g\over 2}n_{0}\right)n'_{2} \label{eq:3.3}
\end{eqnarray}
and
\begin{eqnarray}
\mathcal{E}_{0}^{(4)}&={g\over 2\mathcal{V}^{2}}\sum_{{\bf p},{\bf p}'\neq 0}{gn_{0}\over 2\omega_{\bf p}}{gn_{0}\over 2\omega_{{\bf p}'}}+{gn'_{1}\over 2}{(Mgn_{0})^{3/2}\over 3\pi^{2}\hbar^{3}}+{gn'_{2}\over 2 }{(Mgn_{0})^{3/2}\over 3\pi^{2}\hbar^{3}} \nonumber \\
&+\nu_{12}n'_{1}n'_{2}+g{(Mgn_{0})^{3}\over(3\pi^{2}\hbar^{3})^{2}}. \label{eq:3.4}
\end{eqnarray}
We have employed the fact that at zero temperature, $f_{b\bf p}=0$ and $f_{i{\bf p}}=\theta(p_{iF}-p)$, where $\theta(p)$ is the Heaviside step function and $p_{iF}$ is the Fermi momentum related to the density of unbound fermions $n'_{i}$ by $p_{iF}=(6\pi^{2})^{1/3}{n'_{i}}^{1/3}\hbar$. All the internal energies are assumed to be negative, $\varepsilon_{1}$, $\varepsilon_{2}$, $\varepsilon<0$. When computing $\mathcal{E}_{0}^{(4)}$, we have used the well-known Wick-Bloch-De Dominicis theorem \cite{Wick,Bloch} as well as the explicit form of $\mathcal{U}_{\bf p}$ and $\mathcal{V}_{\bf p}$ including the relations:
$$
\mathcal{V}_{\bf p}=-{gn_{0}\over{{p^{2}/2M}+gn_{0}}+\omega_{{\bf p}}}\,\mathcal{U}_{\bf p}, \quad \mathcal{V}_{\bf p}\mathcal{U}_{\bf p}=-{gn_{0}\over 2\omega_{\bf p}}.
$$
As one can see, Eqs. (\ref{eq:3.3}), (\ref{eq:3.4}) contain two sums in which the expressions have the following asymptotic behavior for large values of momentum:
\begin{eqnarray}
{p^{2}\over 2M}+gn_{0}-\omega_{\bf p}\to {Mg^{2}n_{0}^{2}\over p^{2}}, \quad p\to\infty, \nonumber \\
{gn_{0}\over 2\omega_{\bf p}}\to{Mgn_{0}\over p^{2}}, \quad p\to\infty.  \nonumber
\end{eqnarray}
Therefore, the corresponding integrals diverge at the upper limit since the change to integration gives a factor $p^{2}dp$. As we have noted, the divergences are associated with the fact that we have replaced the function $g({\bf p})$ that characterizes the interaction between the bound states by the coupling constant $g$. Indeed, such a replacement is valid only for small momenta but not for computing high-momentum processes. The difficulty is overcome by renormalization of the coupling constant $g$ in the leading term $\mathcal{E}_{0}^{(0)}$,
\begin{equation}\label{eq:3.5}
g\to g+{g^{2}\over\mathcal{V}}\sum_{{\bf p}\neq 0}{M\over p^{2}}-{g^{3}\over{\mathcal{V}^{2}}}\sum_{{\bf p},{\bf p}'\neq 0}\left[{M\over 2\omega_{\bf
p}p\,'^{2}}+{M\over 2\omega_{{\bf p}'}p^{2}}-{M^{2}\over p^{2}p\,'^{2}}\right].
\end{equation}
Then, the corresponding integrals can be easily computed. The results are
\begin{eqnarray}
\mathcal{E}_{0}^{(2)}&={8\over 15}{gn_{0}\over\pi^{2}\hbar^{3}}(Mgn_{0})^{3/2}+{3\over 10}(6\pi^{2})^{2/3}{\hbar^{2}\over m_{1}}{n'_{1}}^{5/3}-\left(|\varepsilon_{1}|-{g\over 2}n_{0}\right)n'_{1} \nonumber \\
&+{3\over 10}(6\pi^{2})^{2/3}{\hbar^{2}\over m_{2}}{n'_{2}}^{5/3}-\left(|\varepsilon_{2}|-{g\over 2}n_{0}\right)n'_{2} \label{eq:3.6}
\end{eqnarray}
and
\begin{eqnarray}
\mathcal{E}_{0}^{(4)}={11\over 18}{g\over(\pi^{2}\hbar^{3})^{2}}(Mgn_{0})^{3}+{g\over 2} {(Mgn_{0})^{3/2}\over 3\pi^{2}\hbar^{3}}(n'_{1}+n'_{2})
+\nu_{12}n'_{1}n'_{2}. \label{eq:3.7}
\end{eqnarray}
If one ignores the terms associated with fermions and $\mathcal{E}_{0}^{(4)}$, then $\mathcal{E}_{0}\approx\mathcal{E}_{0}^{(0)}+\mathcal{E}_{0}^{(2)}$ is the ground state energy density found by Lee and Yang \cite{Lee-Yang} (see also \cite{Huang}). However, in our case, the condensate density $n_{0}$ is expressed through the total densities of fermionic atoms $n_{1}$ and $n_{2}$. It is evident that the following inequalities
\begin{equation} \label{eq:3.8}
|\mathcal{E}_{0}^{(4)}|\ll |\mathcal{E}_{0}^{(2)}|\ll |\mathcal{E}_{0}^{(0)}|
\end{equation}
represent the conditions of applicability of the theory under consideration. As we see below, they allow one to obtain the numerical estimates of the lower and upper values of $n_{0}$.

At zero temperature, the densities of fermionic atoms, according to Eqs. (\ref{eq:2.32}), (\ref{eq:2.39})-(\ref{eq:2.41}), are given by
\begin{eqnarray}
n_{i}=n_{0}+\tilde{n}+n'_{i}, \label{eq:3.9} \\
\tilde{n}={1\over 3\pi^{2}\hbar^{3}}(Mgn_{0})^{3/2}, \quad n'_{i}={1\over 6\pi^{2}\hbar^{3}}(2m_{i}\tilde{\mu}_{i})^{3/2}, \label{eq:3.9'}
\end{eqnarray}
where $\tilde{n}$ is the density of non-condensate bound states describing the quantum depletion of a condensate and $n'_{i}$ are the densities of unbound fermionic atoms. The chemical potential $\tilde{\mu}_{i}$ coincides with the Fermi energy. From Eqs. (\ref{eq:3.9}) and (\ref{eq:3.9'}), one obtains
\begin{equation}\label{eq:3.10}
\tilde{\mu}_{i}=\varepsilon_{iF}={{(3\pi^{2}\hbar^{3})^{2/3}}\over{2^{1/3}m_{i}}}
\left[n_{i}-n_{0}-{(Mgn_{0})^{3/2}\over 3\pi^{2}\hbar^{3}}\right]^{2/3}.
\end{equation}
Next, Eqs. (\ref{eq:2.22}) allow us to write the basic Eq. (\ref{eq:2.19}) for the condensate density in the form
\begin{eqnarray}
{(3\pi^{2}\hbar^{3})^{2/3}\over {2}^{1/3}m_{1}}&\left[n_{1}-n_{0}-{(Mgn_{0})^{3/2}\over{3\pi^{2}\hbar^{3}}}\right]^{2/3} \nonumber \\ &+{(3\pi^{2}\hbar^{3})^{2/3}\over {2}^{1/3}m_{2}}\left[n_{2}-n_{0}-{(Mgn_{0})^{3/2}\over 3\pi^{2}\hbar^{3} }\right]^{2/3}-\epsilon_{t}=0, \label{eq:3.11}
\end{eqnarray}
where
\begin{equation}\label{eq:3.11'}
\epsilon_{t}=\varepsilon-\varepsilon_{1}-\varepsilon_{2}=|\varepsilon_{1}|+|\varepsilon_{2}|-|\varepsilon|>0.
\end{equation}
Equation (\ref{eq:3.11}) gives the equilibrium condensate density of diatomic bound states at zero temperature and fixed densities $n_{1}$, $n_{2}$ of fermionic atoms. In principle, the bound state energy $\varepsilon$ should be found from the Schr\"{o}dinger equation for two atoms (see Eq. (\ref{eq:1.3})).

In order to analyze Eq. (\ref{eq:3.11}) numerically, we address the experimental data.
The experiments with ultracold atomic gases are usually carried out at densities $n\sim 10^{12}-10^{13}$ cm${}^{-3}$. To be concrete, let us consider a mixture of ${}^{6}$Li and ${}^{173}$Yb atoms with masses $m_{1}\approx 9.985\cdot 10^{-24}$ g and $m_{2}\approx 2.871\cdot 10^{-22}$ g, respectively. The simultaneous quantum degeneracy of such a mixture was experimentally realized in Ref. \cite{Hara}. As has been noted by the authors, it provides a good basis for creating ultracold molecules. In the same work, the absolute value of the $s$-wave scattering length for a collision of ${}^{6}$Li and ${}^{173}$Yb atoms has also been measured, $|a_{12}|\approx 0.9\cdot 10^{-7}$ cm. The mass of the ${}^{6}$Li--${}^{173}$Yb bound states is equal to $M\approx 2.971\cdot 10^{-22}$ g, while the reduced mass is $m_{*}\approx9.65\cdot 10^{-24}$ g. The interaction between the bound states is described by the coupling constant, which, in accordance with Eq. (\ref{eq:2.37'}), has a value $g\approx 1.30\cdot 10^{-37}$ erg$\cdot$ cm${}^{3}$. Note that the positive value of $g$ guarantees the stability of the spectrum given by Eq. (\ref{eq:2.26}).
At the above given  values of physical parameters, Eq. (\ref{eq:3.11}) has a solution when the energy $\epsilon_{t}\sim 10^{-23}-10^{-22}$ erg (see Eq. (\ref{eq:3.11'})). It is worth stressing that this order of magnitude agrees with the typical values of the bound state energies $\varepsilon$ for molecules observed experimentally in ultracold dilute gases \cite{Kohler,Grimm}. Since the absolute value of the bound state energy of some molecules can be much higher, $\varepsilon\sim 10^{-18}$ erg (or $\varepsilon\sim 1$ GHz) \cite{Voigt}, the corresponding fermionic atoms with energies $\varepsilon_{1}$ and $\varepsilon_{2}$, according to Eq. (\ref{eq:3.11'}), are in highly excited states.
\begin{figure}
\centering
\includegraphics[width=0.5\textwidth]{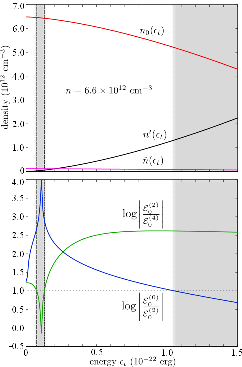}
\caption{Top panel: the dependencies of the condensate density $n_{0}$, density of unbound fermionic atoms $n'$, and non-condensate density of bound states $\tilde{n}$ (quantum depletion) on the parameter $\epsilon_{t}$ at the total density $n=6.6\cdot 10^{12}$ cm${}^{-3}$ (symmetric case). Bottom panel: Extracting the interval of $\epsilon_{t}$ in which the developed perturbative approach is valid (in accordance with Eqs. (\ref{eq:3.8}) and (\ref{eq:3.14})-(\ref{eq:3.16})). The corresponding unshaded regions specify the lower and upper values of the condensate density.}
\end{figure}

{\bf Symmetric case.} Consider the case of equal densities of fermionic atoms, $n_{1}=n_{2}=n$. Then, Eq. (\ref{eq:3.11}) takes the form
\begin{eqnarray}\label{eq:3.12}
{(3\pi^{2}\hbar^{3})^{2/3}\over {2}^{1/3}m_{*}}\left[n-n_{0}-{(Mgn_{0})^{3/2}\over 3\pi^{2}\hbar^{3} }\right]^{2/3}-\epsilon_{t}=0, \quad m_{*}={m_{1}m_{2}\over{m_{1}+m_{2}}}.
\end{eqnarray}
From Eq. (\ref{eq:3.12}), one finds the density of unbound fermionic atoms,
\begin{equation} \label{eq:3.13}
n'={\sqrt{2}(m_{*}\epsilon_{t})^{3/2}\over 3\pi^{2}\hbar^{3}}.
\end{equation}
Note that Eq. (\ref{eq:3.12}) is equivalent to Eqs. (\ref{eq:3.9}) and (\ref{eq:3.9'}) when $n_{1}=n_{2}=n$. Equations (\ref{eq:3.10}) and (\ref{eq:3.12}) give a relationship between the Fermi energies and parameter $\epsilon_{t}$,
$$
\varepsilon_{iF}={m_{*}\over m_{i}}\epsilon_{t}.
$$
The ground state energy density is given by Eq. (\ref{eq:3.1}), in which
\begin{eqnarray}
\mathcal{E}_{0}^{(0)}=-|\varepsilon| n_{0}+{gn_{0}^{2}\over 2}, \label{eq:3.14} \\
\mathcal{E}_{0}^{(2)}={8\over 15}{gn_{0}\over \pi^{2}\hbar^{3}}(Mgn_{0})^{3/2}+{3\over 10}(6\pi^{2})^{2/3}{\hbar^{2}\over m_{*}}{n'}^{5/3}-(|\varepsilon_{1}|+|\varepsilon_{2}|-gn_{0})n', \label{eq:3.15} \\
\mathcal{E}_{0}^{(4)}={11\over 18}{g\over(\pi^{2}\hbar^{3})^{2}}(Mgn_{0})^{3}+{gn'\over 3\pi^{2}\hbar^{3}}(Mgn_{0})^{3/2}+\nu_{12}{n'}^{2}. \label{eq:3.16}
\end{eqnarray}

Now we present the numerical analysis of Eq. (\ref{eq:3.12}), (\ref{eq:3.14})-(\ref{eq:3.16}) at the above given values of the physical parameters. The top panel of Fig. 1 shows the dependencies of the condensate density $n_{0}$, density of unbound fermionic atoms $n'$, and non-condensate density $\tilde{n}$ (the quantum depletion of a condensate) on $\epsilon_{t}$ at the fixed total density $n=6.6\cdot 10^{12}$ cm${}^{-3}$. All these quantities enter the ground state energy and, therefore, their values must be such that they satisfy the inequalities given by Eq. (\ref{eq:3.8}). As mentioned, these inequalities define the boundaries of the developed perturbative approach and allow one to estimate the lower and upper values of the condensate density $n_{0}$ at the given total density of fermionic atoms. In order to solve such a problem, we plot the dependencies of $\log\left|\mathcal{E}_{0}^{(0)}/\mathcal{E}_{0}^{(2)}\right|$ and $\log\left|\mathcal{E}_{0}^{(2)}/\mathcal{E}_{0}^{(4)}\right|$ on $\epsilon_{t}$ in the bottom panel. They are plotted at the same total density $n$ and in the same interval of $\epsilon_{t}$ as in the top panel. We assume that the inequalities are {\it a fortiori} satisfied when $\left|\mathcal{E}_{0}^{(0)}\right|$ and $\left|\mathcal{E}_{0}^{(2)}\right|$ as well as $\left|\mathcal{E}_{0}^{(2)}\right|$  and $\left|\mathcal{E}_{0}^{(4)}\right|$ differ at least by one order of magnitude, so that $\left|\mathcal{E}_{0}^{(0)}/\mathcal{E}_{0}^{(2)}\right|\sim\left|\mathcal{E}_{0}^{(2)}/\mathcal{E}_{0}^{(4)}\right| \sim 10$ (the same assumption is used below when analyzing the asymmetric case, see Fig. 2). Therefore, the inequalities given by Eq. (\ref{eq:3.8}) are true only in those regions where both curves remain above unity. At the given total density, the unshaded regions indicate the intervals of $\epsilon_{t}$ in which the developed theory is valid and specifies the lower and upper values of the condensate density $n_{0}$. For the above value of $n$, $\epsilon_{t}$ varies from $0$ erg to $0.07\cdot 10^{-22}$ erg and from $0.14\cdot 10^{-22}$ erg to $1.05\cdot 10^{-22}$ erg. In these intervals of energy, the condensate density changes from $6.48\cdot 10^{12}$ cm${}^{-3}$ to $6.46\cdot 10^{12}$ cm${}^{-3}$ and from $6.42\cdot 10^{12}$ cm${}^{-3}$ to $5.2\cdot 10^{12}$ cm${}^{-3}$. As for the shaded regions, they indicate the intervals of $\epsilon_{t}$, where the developed theory is not valid and, consequently, we cannot say if there is a condensate in the corresponding interval of energies. When plotting the curves, we assume (see also Fig. 2), for simplicity, that all the internal energies are of the same order of magnitude, $|\varepsilon_{1}|\sim|\varepsilon_{2}|\sim|\varepsilon|\sim|\epsilon_{t}|$.

\begin{figure}
\centering
\includegraphics[width=0.97\textwidth]{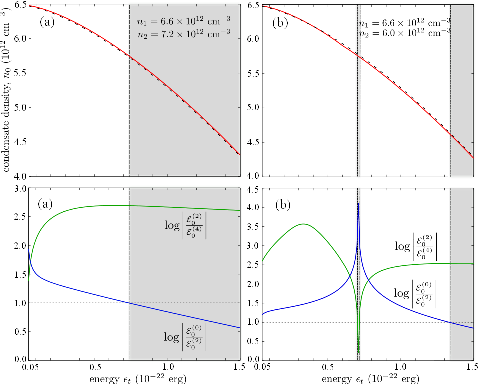}
\caption{Top panels: the dependence of the condensate density $n_{0}$ on the parameter $\epsilon_{t}$: (a) the dominance of the heavy atomic component, $n_{1}<n_{2}$; (b) the dominance of the light atomic component, $n_{2}<n_{1}$. Lower panels: Extracting the intervals of $\epsilon_{t}$ in which the perturbative approach is valid (in accordance with Eqs. (\ref{eq:3.2}) and (\ref{eq:3.6})-(\ref{eq:3.8})): (a) the dominance of the heavy atomic component, $n_{1}<n_{2}$; (b) the dominance of the light atomic component, $n_{2}<n_{1}$. The corresponding unshaded regions specify the lower and upper values of the condensate density.}
\end{figure}

{\bf Asymmetric case.} Consider now a more realistic situation from the experimental point of view, when the densities of fermionic atoms do not coincide, $n_{1}\neq n_{2}$. To analyze it, we need to use the more general Eq. (\ref{eq:3.11}) and Eqs. (\ref{eq:3.2}), (\ref{eq:3.6}), (\ref{eq:3.7}) for the ground state energy density. Moreover, two distinct cases are possible: (a) the heavy fermions with masses $m_{2}$ dominate over the light atoms with masses $m_{1}$, so that $n_{2}>n_{1}$ and (b) the density of light fermionic atoms higher than the density of heavy fermions, $n_{1}>n_{2}$.

We start from the case (a) and take the densities equal to $n_{1}=6.6\cdot 10^{12}$ cm${}^{-3}$ and $n_{2}=7.2\cdot 10^{12}$ cm${}^{-3}$ (the dominance of heavy atoms is of the order of 10\%). Then the results of numerical analysis are presented in the left two panels of Fig. 2. The solid line in the top panel demonstrates the condensate density as a function of $\epsilon_{t}$. The meaning of the dotted line is explained below. The lower left panel shows the dependencies of $\log\left|\mathcal{E}_{0}^{(0)}/\mathcal{E}_{0}^{(2)}\right|)$ and $\log\left|\mathcal{E}_{0}^{(2)}/\mathcal{E}_{0}^{(4)}\right|$ on $\epsilon_{t}$, which allow one to extract, absolutely in the same manner as described in Fig. 1, the interval of $\epsilon_{t}$ in which the developed perturbative approach is valid. The unshaded region corresponding to this interval specifies the upper and lower values of the condensate density, $5.76\cdot 10^{12}$\,cm${}^{-3}$$\lesssim n_{0}\lesssim 6.48\cdot 10^{12}$\,cm${}^{-3}$. If the density of heavy atoms increases at the fixed density of light atoms, then the region becomes more narrow.

Two right panels of Fig. 2 describe the opposite case (b). The densities are taken to be $n_{1}=6.6\cdot 10^{12}$ cm${}^{-3}$ and $n_{2}=6.0\cdot 10^{12}$ cm${}^{-3}$ so that the dominance of light fermionic atoms over the heavy atoms also makes up 10\%. In this case, the dependence of the condensate density $n_{0}$ on $\epsilon_{t}$ is shown on the top right panel. A similar analysis, as has been done above, enables us to conclude that there exist two intervals of $\epsilon_{t}$ which define the boundaries of the developed theory. As in Fig. 1, the unshaded regions, corresponding to these intervals, specify the possible values of the condensate density. In the left domain, $n_{0}$ varies from $6.48\cdot 10^{12}$ cm${}^{-3}$ to $5.76\cdot 10^{12}$ cm${}^{-3}$ and in the right, from $5.73\cdot 10^{12}$ cm${}^{-3}$ to $4.63\cdot 10^{12}$ cm${}^{-3}$. In these regions $\epsilon_{t}$ changes in the intervals $0.05\cdot 10^{-22}$ erg $\lesssim\epsilon_{t}\lesssim 0.7\cdot 10^{-22}$ erg and $0.72\cdot 10^{-22}$ erg $\lesssim\epsilon_{t}\lesssim 1.33\cdot 10^{-22}$ erg, respectively. The shaded regions show the interval of energies in which the perturbative approach becomes inapplicable.

To compare both cases (a) and (b), we plot the dotted lines in each top panel of Fig. 2. These lines correspond to the symmetric case, when $n_{1}=n_{2}=6.6\cdot 10^{12}$ cm${}^{-3}$ (see Fig. 1). Therefore, Fig. 2(a) shows that the dominance of the heavy atomic component over the light component ($n_{1}<n_{2}$) increases the condensate fraction of heteronuclear bound states, while the large values of $\varepsilon_{t}$ suppress the condensate. As shown in Fig. 2(b), the dominance of the light atomic component ($n_{2}<n_{1}$) may result in increasing or decreasing the condensate density $n_{0}$ depending on the values of $\epsilon_{t}$.

Finally, it is worth noting that the characteristic values of the physical parameters presented above are such that the following inequality is satisfied:
$$
gn_{0}\ll \epsilon_{t}.
$$
This shows that the interaction should be weak enough to ensure the existence of the 'stable' bound states. The decay processes can be studied on the basis of consistent derivation of kinetic equation for bound states of particles (see, e.g., \cite{Peletm}).

\section{Conclusion}

We proposed a microscopic approach for studying a many-body system of interacting fermionic atoms of two species that are in thermodynamic equilibrium with their condensed Fermi-Fermi heteronuclear bound states. Both this approach and the formulated problem itself dealing with Bose-Einstein condensation of heteronuclear bound states formed in a Fermi gas of two atomic species seem to be novel. A starting point for its solution is the microscopic Hamiltonian \cite{Pelet-Slyus} in which the interactions of bound states with themselves and with fermionic atoms are expressed through the interaction amplitudes of unbound fermions. This fact allows one to apply the Bogoliubov approach if the components of the system weakly interact. The applicability condition for the second quantized Hamiltonian and approximate second quantization method itself is the smallness of the average kinetic energy (or temperature) compared to the energies, which specify the spectrum of the bound states. We have found the single-particle excitation spectrum in the framework of modified Bogoliubov's approach applied to the mentioned Hamiltonian.  Outwardly, it looks like the well-known spectrum for a weakly interacting Bose gas with a condensate. However, its specific feature is that the condensate density of the bound states in the spectrum is expressed through the total densities of fermionic atoms of both species. The ground state energy of the system as well as the equation for the condensate density at zero temperature were also obtained. The latter enables one to study any two-component Fermi mixture coexisting in equilibrium with heteronuclear bound states formed from the same fermions. To be specific and to analyze the equation numerically, we have considered a mixture of fermionic ${}^{6}$Li and ${}^{173}$Yb atoms. The simultaneous quantum degeneracy in this mixture was realized experimentally in \cite{Hara}, where the authors claimed that this mixture provides a good basis for the creation of ultracold heteronuclear molecules. Our analysis shows, in particular, that by changing the ratio between the densities of light and heavy atoms it is possible to achieve both increasing and decreasing condensate density. We analyzed the applicability conditions for the developed theory, which allow one to estimate the lower and upper values for the condensate density at the fixed total densities of fermionic atoms. The obtained results could be useful in efforts to discover experimentally the condensation of heteronuclear molecules formed in a mixture of two-species Fermi gas.

The following problems seem to be interesting here: to examine the role of temperature effects on the density of condensed bound states and to investigate, on the basis of the used Hamiltonian, the coexistence of molecular condensate with a superfluid Fermi gas \cite{Yatsenko}. As for the first problem, the simplest way to solve it is to take into account the temperature terms in Eqs. (\ref{eq:2.39})-(\ref{eq:2.41}) if both components of a Fermi gas are degenerate. However, more exotic situations may also be realized when both components of a Fermi gas are non-degenerate or only one of them is degenerate. These problems form the current research work of the authors.

\ack
The authors would like to thank A.G. Sotnikov and M.V. Bondarenco for useful discussions.

\end{document}